\documentclass[sigconf]{acmart}

\copyrightyear{2026}
\acmYear{2026}
\setcopyright{cc}
\setcctype{by}
\acmConference[UMAP '26]{34th ACM Conference on User Modeling, Adaptation and Personalization}{June 08--11, 2026}{Gothenburg, Sweden}
\acmBooktitle{34th ACM Conference on User Modeling, Adaptation and Personalization (UMAP '26), June 08--11, 2026, Gothenburg, Sweden}
\acmDOI{10.1145/3774935.3806657}
\acmISBN{979-8-4007-2311-7/2026/06}

\usepackage{float}
\usepackage{subcaption}
\usepackage{enumitem}
\usepackage{hyperref}

\usepackage{booktabs}
\usepackage{siunitx}
\usepackage[noabbrev,nameinlink]{cleveref}
\crefformat{section}{\S#2#1#3} %
\crefformat{subsection}{\S#2#1#3}
\crefformat{subsubsection}{\S#2#1#3}
\crefrangeformat{section}{\S#3#1#4 to~#5#2#6}
\crefrangeformat{subsection}{\S#3#1#4 to~#5#2#6}
\crefmultiformat{section}{\S(#2#1#3)}{ and~#2#1#3}{,#2#1#3}{ and~#2#1#3}
\crefmultiformat{subsection}{\S#2#1#3}{ and~#2#1#3}{, #2#1#3}{ and~#2#1#3}

\crefname{equation}{Eq.}{Eqs.}
\Crefname{equation}{Equation}{Equations}

\usepackage{url}

\usepackage{graphicx}

\newcommand{\note}[2][\relax]{%
    \textcolor{BurntOrange}{#2}%
    \ifx#1\relax %
    \else
        \textcolor{gray}{\footnotesize~[note: #1]}%
    \fi
}

\newcommand{\subsubsub}[1]{\textbf{\textit{#1.}}}

\newcommand{\smarrow}[1]{\scalebox{0.8}{$#1$}}

\usepackage{xspace}
\newcommand{\ocd}{\textsc{OCDegen}\xspace}

\newcommand{\blockies}{\textsc{Blockies}\xspace}
\newcommand{\twofourtwo}{\textsc{Two4Two}\xspace}

\graphicspath{{figures/}}

\begin{document}

\title[Raising the Stakes]
{Raising the Stakes: Assessing the Influence of Stakes on User Reliance Behavior in Human-AI Decision-Making}

\author{David S. Johnson}
\email{djohnson@techfak.uni-bielefeld.de}
\orcid{1234-5678-9012}
\affiliation{%
  \institution{Bielefeld University}
  \city{Bielefeld}
  \country{Germany}
}

\renewcommand{\shortauthors}{Johnson}

\begin{abstract}
Human-AI collaboration is often proposed to improve high-stakes decision-making, yet the influence of increased stakes and imperfect AI on decision-making strategies is not fully understood. Studying such behavior in realistic settings is challenging, as application-grounded evaluations are costly, rely on experts, or lack meaningful consequences for decision errors. To address this, we introduce \blockies, a parametric dataset generator for visual diagnostic tasks, and conduct an empirical study examining how perceived stakes influence reliance calibration and behavior. Results show that raised stakes lead to longer deliberation, but less calibrated reliance, with participants increasingly deferring to incorrect AI advice as decision time increased. These findings highlight that increased effort under higher stakes does not necessarily improve reliance calibration and show the importance of accounting for stakes when evaluating human-AI decision-making.
\end{abstract}




\maketitle

\section{Introduction}\label{sec:intro}

Artificial Intelligence (AI) is increasingly being used in high-stakes domains, such as medicine~\cite{chanda_dermatologist-like_2024,sutton_overview_2020,lekadir_future-ai_2025} and affective computing~\cite{lutz_prospective_2022,saakyan_scalable_2023,johnson_explainable_2024}, with the aim of enhancing decision-making through Human-AI collaboration~\cite{lai_towards_2023}. 
However, a recent meta-analysis~\cite{vaccaro_when_2024} suggests that these benefits have yet to be fully realized, as Human-AI teams often perform worse than either humans or AI alone. 

To address this, researchers have explored various interventions, including explainable AI (XAI) approaches that provide decision rationales~\cite{antoniadi_current_2021} or novel interaction paradigms that aim to counteract cognitive biases~\cite{bucinca_trust_2021, miller_explainable_2023, le_towards_2024}. Research has shown that explanations may also reduce human-AI performance by inducing mistrust in the AI~\cite{amarasinghe_importance_2024,chen_understanding_2023}, through both overreliance~\cite{chen_understanding_2023,bucinca_trust_2021} and underreliance~\cite{vasconcelos_explanations_2023}.
Alternative interventions such as selecting appropriate explanation types~\cite{chen_understanding_2023}, applying cognitive-forcing techniques~\cite{bucinca_trust_2021}, or allowing users to explore their own hypotheses~\cite{miller_explainable_2023, le_towards_2024} have shown promise to address these issues. However, research in Human-AI decision-making is still in its early stages and would benefit from more empirical studies~\cite{lai_towards_2023,suh_fewer_2025} to understand how different interventions influence users' behavioral patterns and reliance strategies.

A significant challenge in advancing Human-AI decision-making is that current evaluations rely on proxy tasks~\cite{vasconcelos_explanations_2023, bucinca_trust_2021, hase_evaluating_2020,swaroop_personalising_2025}, or small interpretable datasets~\cite{chen_understanding_2023, le_towards_2024,alufaisan_does_2020}, to elicit decision-making behavior while keeping tasks easy to learn and scalable. Such studies, however, limit task complexity and may not accurately predict real-world outcomes~\cite{bucinca_proxy_2020,amarasinghe_importance_2024}.

Application-grounded evaluation~\cite{doshi-velez_towards_2017} that involves real tasks and experts is a gold standard~\cite{gambetti_survey_2026,amarasinghe_importance_2024}, but it is often inaccessible to researchers or too costly and time-consuming to scale effectively~\cite{amarasinghe_importance_2024, chanda_dermatologist-like_2024, zytek_sibyl_2022}. 
Accessible, online application-grounded studies have been conducted~\cite{humer_reassuring_2024, morrison_evaluating_2023, sixt_users_2022}, but their assessments were heterogeneous, limiting reproducibility and cross-study comparison. Furthermore, studies rarely model the \textit{perceived stakes} of decisions, and those that employ high-stakes tasks~\cite{humer_reassuring_2024, chanda_dermatologist-like_2024} lack real or perceived consequences for participants. Such consequences, however, may influence how a user chooses to rely on an AI~\cite{vasconcelos_explanations_2023, starcke_decision_2012, Wang2022_}. 

To address these challenges, we introduce \blockies\footnote{OSF Repository: \url{https://tinyurl.com/umap26-supplementary}}, a parametric dataset generator for visual diagnostic tasks that provides fine-grained control over task complexity, bias, and ambiguity suitable for online studies with general users. Furthermore, we conducted an empirical study, based on \blockies, that incorporates risk framing and monetary incentives to modulate perceived stakes. Findings from the study suggest that raised stakes led to longer decision-making times that did not improve decision quality, as participants showed increased overreliance with longer deliberation.

\section{\blockies Datasets}

\begin{figure}[t]
    \centering
    \resizebox{0.90\linewidth}{!}{
        \begin{minipage}{\linewidth}
            \centering
            \begin{subfigure}[t]{0.48\linewidth}
                \centering
                \includegraphics[width=\linewidth]{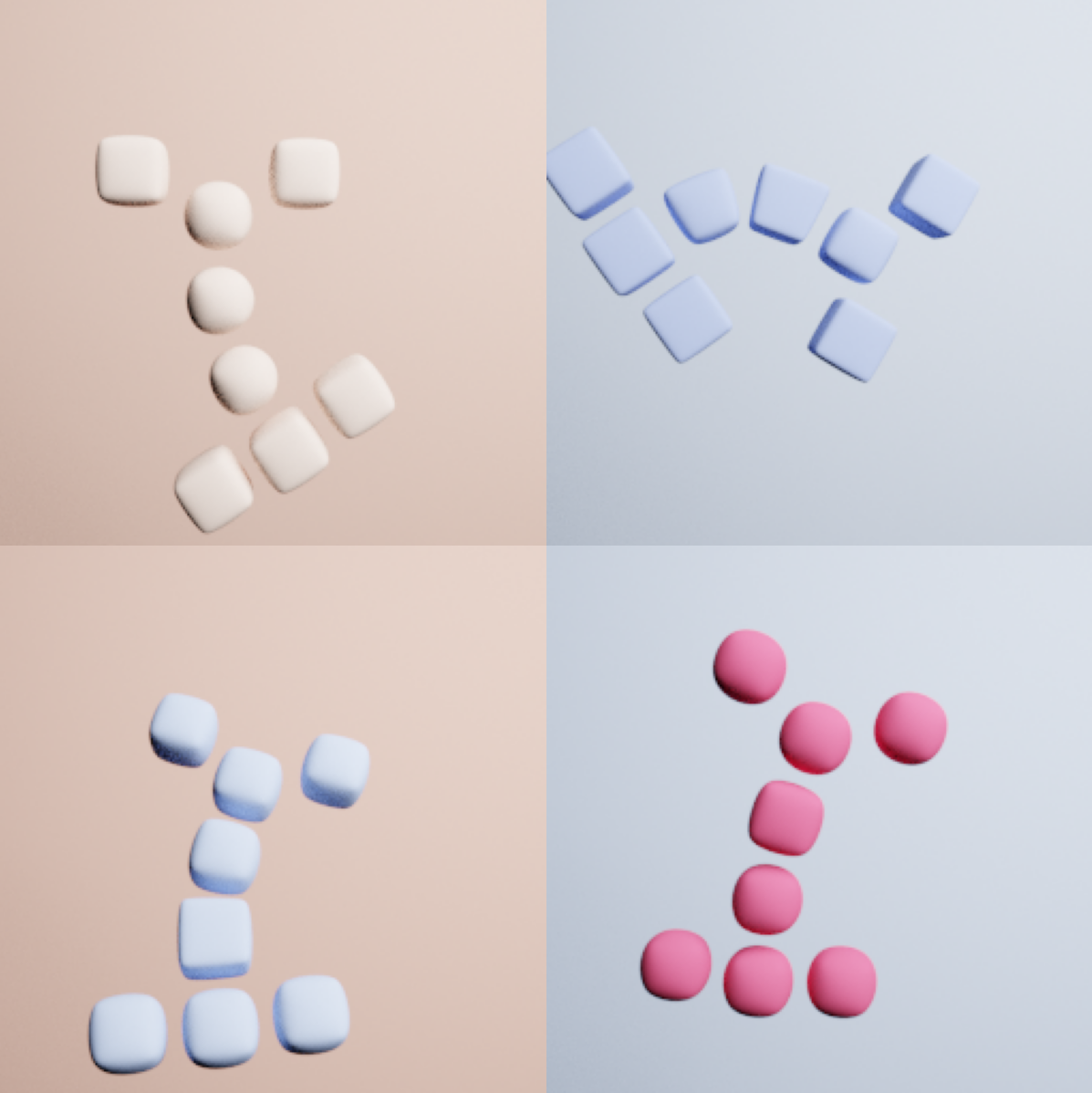}
                \caption{Healthy \blockies}
                \label{fig:blockies:healthy}
            \end{subfigure}
            \hfill
            \begin{subfigure}[t]{0.48\linewidth}
                \centering
                \includegraphics[width=\linewidth]{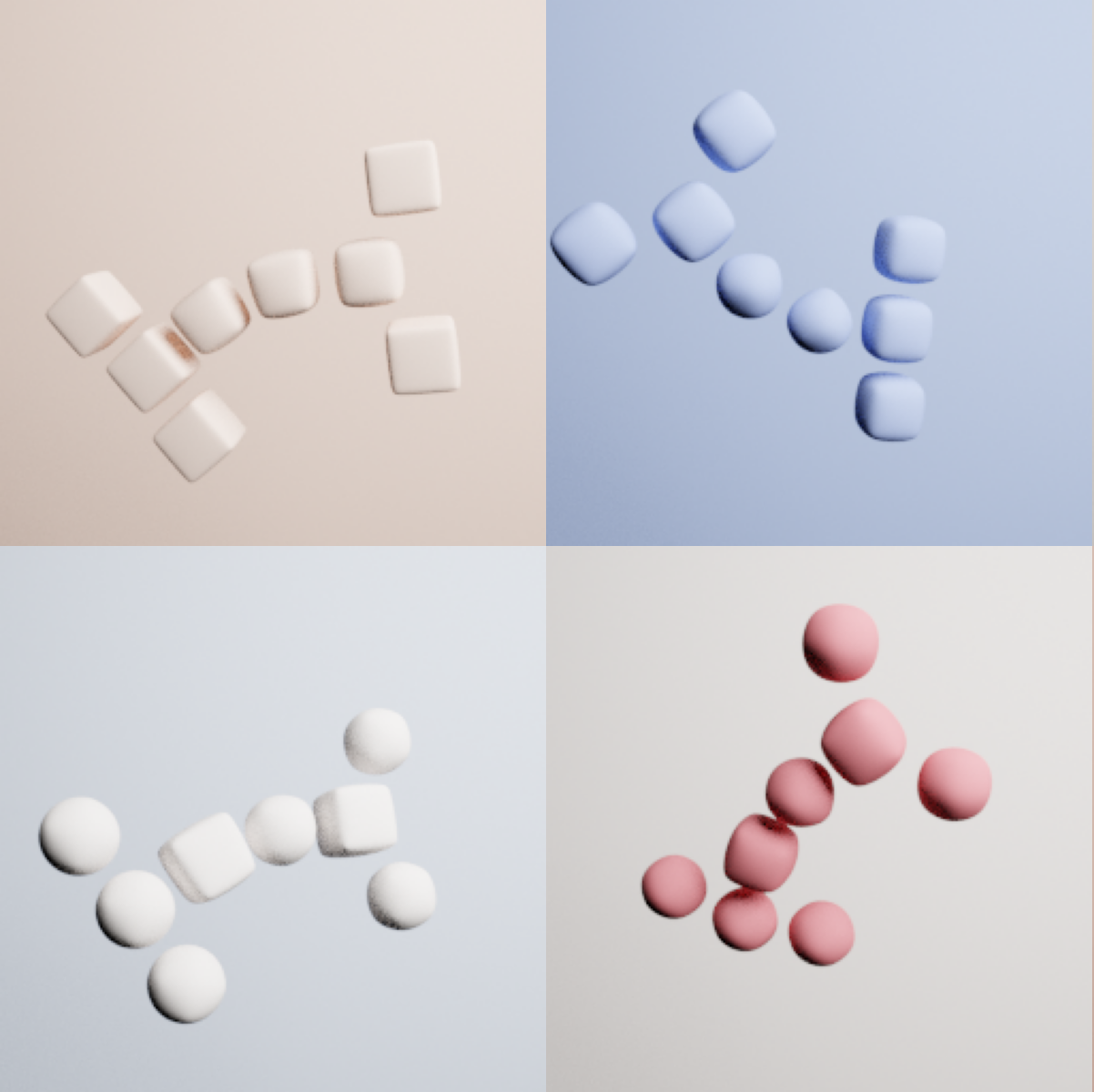}
                \caption{OCDegen \blockies}
                \label{fig:blockies:ocd}
            \end{subfigure}
        \end{minipage}
    }
    \caption{Sample \textit{Blocky} instances generated with \blockies.}
    \label{fig:blockies}
    \Description{Eight example Blocky X-ray images generated by the Blockies dataset generator, showing virtual creatures affected by Osteocuboid Degeneration with visual abnormalities characteristic of the condition.}
\end{figure}

\textsc{Blockies} is a parametric dataset generator\footnote{Code: \url{https://github.com/davidsjohnson/blockies-haic}} for designing visual diagnostic tasks in which users diagnose virtual 3D creatures, called \textit{Blockies}, based on observed traits and symptoms. Each \textit{Blocky} consists of eight blocks (or bones), and has traits, such as bone shape, posture, and color, that can be systematically manipulated for fine-grained control of dataset biases. Each trait is governed by a sampling distribution, and sampled values are used to parametrically generate a 3D scene that is rendered as an image.

\blockies extends the \twofourtwo data generator~\cite{sixt_users_2022}, but rather than relying on a single trait for class discrimination (e.g., the leg position as in \twofourtwo), \blockies diagnostic tasks are defined by assigning sets of symptoms to different diagnostic classes. Each symptom has a unique sampling distribution that is distinct from its corresponding trait. 

For this study, we designed a diagnosis task inspired by common diagnostic practices~\cite{chanda_dermatologist-like_2024}. Participants diagnose a condition we term \textit{Osteocuboid Degeneration} (\ocd) by reviewing so-called X-rays. \ocd is defined by the presence of at least two assigned symptoms, whereas healthy cases exhibit fewer than two, see \Cref{fig:blockies} for examples\footnote{\href{https://tinyurl.com/umap26-supplementary}{See the tutorials in the supplement} for a complete list of traits and symptoms.}.

Two main datasets were generated: \textit{development} and \textit{evaluation}. \textit{Development} consists of training, validation, and testing sets for model development. \textit{Evaluation} has a single evaluation set. Each dataset is labeled with the two diagnostic classes (healthy and \ocd). All X-rays in \textit{development} were generated using a common set of trait sampling distributions. X-rays in \textit{evaluation} were generated with broader distributions for positional parameters (pitch, yaw, and roll), resulting in greater variability in body position, while keeping the distributions of all other traits consistent with the \textit{development} data, simulating domain shift~\cite{gretton_covariate_2009}.

MobileNetV2~\cite{sandler_mobilenetv2_2019}, trained on \textit{development}, is the basis for generating AI recommendations. The model achieved \SI{90.43}{\percent} accuracy on the \textit{development} test data, and performance dropped to \SI{69.37}{\percent} on the \textit{evaluation} test set. This performance decrease introduces variability in predictions and decisions.

\section{User Study}\label{sec:userstudy}

We conducted an empirical study using \blockies to study the impact of raised stakes on reliance behavior in human-AI decision-making. The university's Ethics Board approved the study. We preregistered\footnote{OSF Registration: \url{https://tinyurl.com/umap26-preregistration}} the analysis of performance metrics to reduce the risk of selective reporting, enhance the validity of hypothesis testing, and support reproducibility~\cite{nosek_preregistration_2017}.

\subsection{Hypotheses}

Research suggests that raised stakes may lead to more calibrated reliance in human-AI collaboration by increasing cognitive motivation~\cite{vasconcelos_explanations_2023, Wang2022_}, which has been shown to reduce overreliance and improve error detection with a more critical AI evaluation~\cite{bucinca_trust_2021}. Such scrutiny may also reduce trust in AI, particularly when users detect errors~\cite{Dietvorst2015_}. Therefore, we posed the following hypotheses:

\begin{enumerate}[label=\textbf{H\arabic*}, leftmargin=*]
    \item\label{h1} Diagnosis accuracy will be higher in the raised stakes group than in the low-stakes group.
    \item\label{h2} Overreliance will be lower in the raised stakes group than in the low-stakes group.
    \item\label{h5} Participants in the raised stakes group will take more time to make diagnoses.
    \item\label{h3} Participants in the raised stakes group will have lower subjective trust in the AI. 
\end{enumerate}

\subsection{User Study Design}

The study followed a between-subjects design, with participants randomly assigned (balanced by gender) to either the low- or raised-stakes condition. 

\subsubsub{Task Procedure}
Participants completed two diagnosis rounds: a baseline without AI and an AI-supported phase with diagnosis recommendations. The same X-rays were used in both rounds, but presented in different orders. The first six samples for each participant were always correct AI predictions to limit confounds from anchoring bias~\cite{nourani_anchoring_2021, Dietvorst2015_}.  Pilot studies indicated that participants were unaware of the repeated samples.

\subsubsub{Study Datasets}
Two datasets were created for the study: a tutorial set (20 X-rays; 14 correct, 6 incorrect AI predictions) and an evaluation set (40 X-rays; 28 correct, 12 incorrect), sampled from the development and evaluation test sets, respectively.

\subsubsub{Stakes Modulation}
We conceptualize stakes as perceived consequences of decision outcomes through performance-based monetary incentives and risk framing. In the low-stakes condition, participants were informed they would receive a \SI{1.50}{\pounds} bonus, which would be reduced if accuracy fell below \SI{75}{\percent} (prorated down to \SI{60}{\percent}). In the raised-stakes condition, participants were informed they would earn \SI{4.50}{\pounds}, but risked losing the bonus if accuracy fell below \SI{90}{\percent} (prorated down to \SI{75}{\percent}). To ensure fairness, all participants ultimately received the full \SI{4.50}{\pounds} bonus.

Risk framing further reinforced the stakes. In the raised-stakes condition, \ocd was presented as life-threatening with invasive treatment, accompanied by a warning in the interface, whereas in the low-stakes condition it was described as mild with low-risk treatment and no warning.

\begin{figure*}[t]
    \centering
    \begin{subfigure}[t]{0.717\linewidth}
        \centering
        \includegraphics[width=0.98\linewidth]{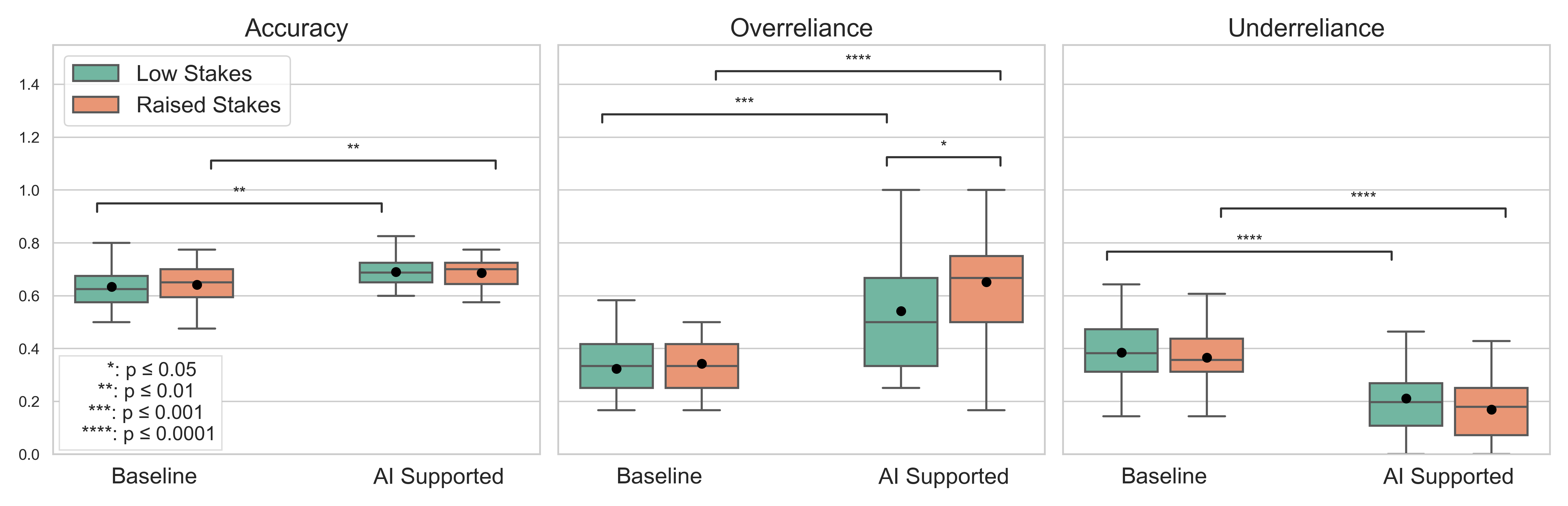}
        \label{fig:hai-perf:perf}
    \end{subfigure}
    \hfill
    \begin{subfigure}[t]{0.277\linewidth}
        \centering
        \includegraphics[width=0.98\linewidth]{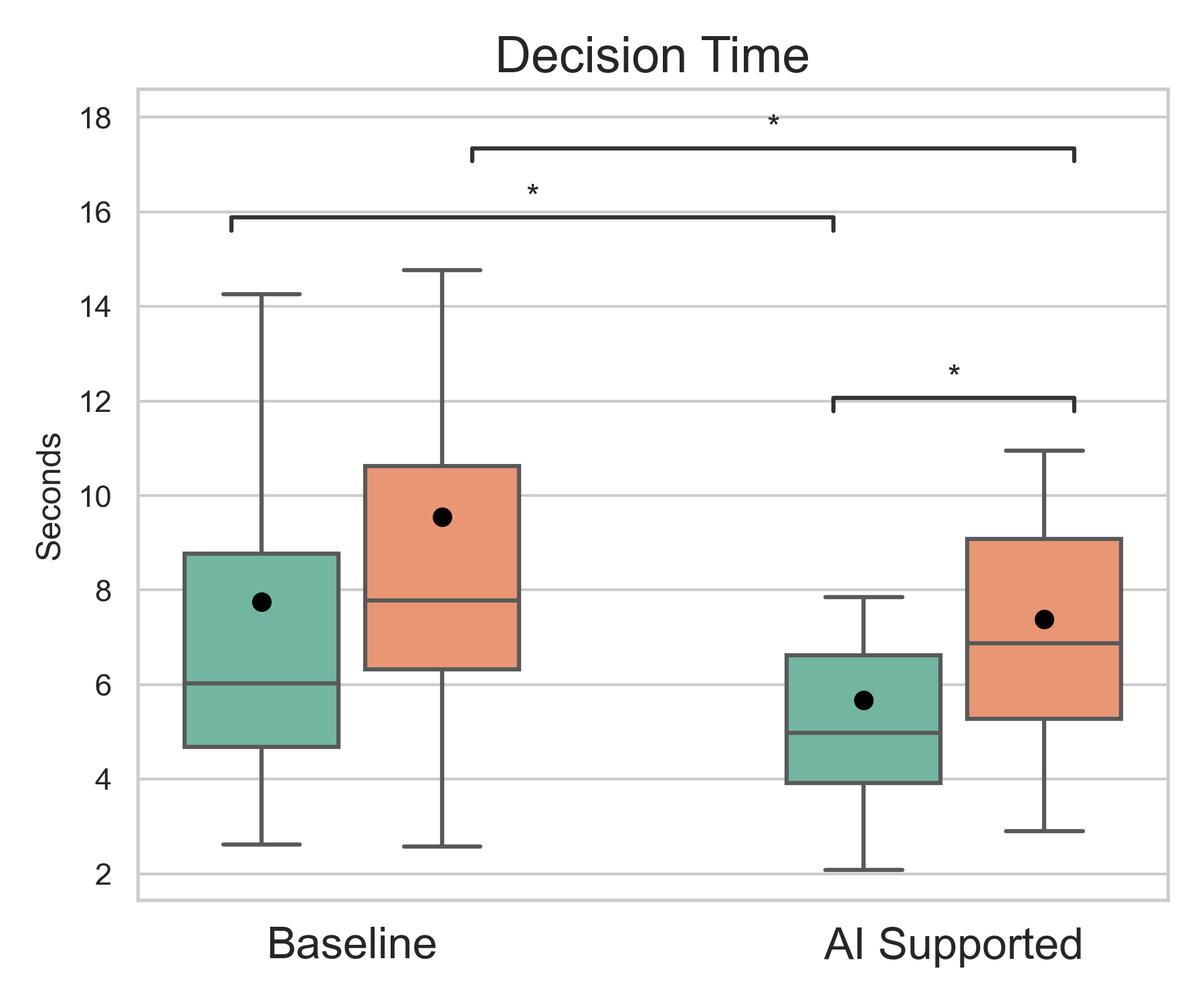}
    \end{subfigure}
    \caption{Analysis of Human-AI performance metrics and decision time. \textit{Baseline} shows results without AI, and \textit{AI Supported} with AI recommendations. Only significant differences ($p\leq0.05$) are shown.}
    \Description{Four side-by-side box plots comparing participant performance and reliance metrics across conditions. From left to right, the panels show Accuracy, AI Alignment, Overreliance, and Underreliance. Each panel contrasts Baseline and AI-Supported conditions, with separate distributions for the low- and raised-stakes groups. Boxes indicate the interquartile range, horizontal lines show the median, and black dots mark the mean. Brackets with asterisks above selected pairs denote statistically significant differences between conditions. A legend identifies the stakes conditions, and a key indicates significance levels.}
    \label{fig:hai-perf}
\end{figure*}

\subsubsub{Study Implementation}
Participants were recruited through the crowdsourcing platform Prolific and offered compensation of \SI{5}{\pounds} plus the performance-based bonus.
After providing informed consent, participants completed questionnaires on demographics, ML knowledge, and the Propensity to Trust scale~\cite{merritt_i_2013}. 

They were then introduced to the \ocd diagnosis task and the bonus structure associated with their assigned condition. Participants completed a tutorial describing the diagnosis of \blockies, including example X-rays with annotated symptoms, and could download the tutorial for reference. Afterwards, they practiced the task using the diagnosis user interface, where AI suggestions and ground truth diagnoses, including contributing symptoms, were provided to familiarize them with the task and interface.

Participants then completed the diagnosis task, followed by a post-task trust questionnaire~\cite{hoffman_measures_2023}.  
Finally, they were debriefed on deception regarding the bonus and reconfirmed consent.

\subsubsub{Metrics}
Participant performance is evaluated using three metrics: \texttt{accuracy}, the proportion of correct final diagnoses; \texttt{overreliance}, the proportion of incorrect human decisions among the incorrect AI diagnoses; and \texttt{underreliance}, the proportion of incorrect human decisions among the correct AI diagnoses.

To further study reliance dynamics across study conditions, we adopt the Appropriateness of Reliance (AoR) framework~\cite{schemmer_appropriate_2023}. AoR focuses specifically on cases where participants' initial decisions differ from the AI, enabling more precise analysis of reliance. We compute relative AI reliance (RAIR), defined as the proportion of correct AI reliance (CAIR) when the user is initially incorrect and the AI is correct, and relative Self Reliance (RSR), defined as the proportion of correct self reliance (CSR) when the participant is initially correct and the AI is incorrect. Together, RAIR and RSR characterize participant-level reliance behavior, with higher values indicating more calibrated reliance. 

\subsection{Data Analysis}
\label{sec:userstudy:analysis}
First, we analyzed the performance metrics, comparing baseline and AI-supported phases as well as low- and raised-stakes conditions using appropriate (depending on normality) paired and independent tests (two-sided, $\alpha = 0.05$), and, to further examine reliance behavior, RAIR and RSR across stakes levels using independent-sample tests (two-sided, $\alpha = 0.05$). Next, we examined the effects of decision time on reliance behavior. Following the AoR framework, we restricted the analysis to cases where participants' initial predictions differed from the AI. We separately analyzed AI-correct and AI-incorrect cases, coding individual decisions as CAIR vs. non-CAIR and CSR vs. non-CSR, respectively. We then modeled the probability of each reliance outcome (CAIR, CSR) as a function of z-standardized decision time, stakes, and their interaction using Bayesian mixed-effects logistic regression models~\cite{Capretto2022}:
{
\setlength{\abovedisplayskip}{6pt}
\setlength{\belowdisplayskip}{6pt}
\setlength{\abovedisplayshortskip}{6pt}
\setlength{\belowdisplayshortskip}{6pt}
\begin{equation}
\begin{split}
p(y = 1) \sim\; 
\text{time\_taken}_i * \text{stakes\_level} \\
\quad + (1 \mid \text{part\_id})
+ (1 \mid \text{blocky\_id})
\end{split}
\label{eq:fem}
\end{equation}
}
where $y \in \{\text{CAIR}, \text{CSR}\}$.

\section{Results}

Participants (\textit{N}=\num{60}) were recruited via Prolific. 
After excluding seven participants for low engagement and one due to corrupted data, \textit{N}=\num{52} participants remained ($n=24$ low-stakes, $n=28$ raised stakes). 
Groups were comparable in demographics, self-reported ML knowledge, and propensity to trust AI.

\subsection{Influence of Stakes on Performance}

We analysed the performance metrics across the baseline and AI-supported phases and between the raised- and low-stakes conditions. The data did not meet normality assumptions, so we employed a two-sided Wilcoxon signed-rank test to compare participant performance across phases, and a two-sided Mann-Whitney U test to evaluate differences between the groups.
 
All metrics showed significant differences ($p \leq 0.05$) between the baseline and AI-support, with strong effect sizes ($|r| > 0.6$), indicating the AI had an impact on decision behavior (\Cref{fig:hai-perf}, left). However, only \texttt{overreliance} resulted in a significant difference between groups during AI support, with a small-medium effect ($r = 0.27$).
These results contradict \ref{h1}, since the raised stakes group did not outperform the low-stakes one in either phase. \ref{h2} is also contradicted as participants tended to overrely more on AI recommendations under raised stakes.

\subsection{Subjective Assessment and Decision Effort}

Decision time was calculated as the average time participants took to make a diagnosis. A Welch's t-test was used to compare decision time across groups, and a paired t-test to compare across phases.  
Decision times for both groups were significantly reduced  ($p \leq 0.05$) in the AI support phase, with a medium effect ($d > 0.44$) (\Cref{fig:hai-perf}, right).  Furthermore, the raised stakes group took significantly longer to make diagnoses with AI support than the low-stakes group ($p \leq 0.05$), with a medium effect ($d=0.44$), supporting \ref{h5}.

Subjective assessment of the task was measured using participants' post-task scores on the Trust in XAI scale, and through participants' responses to a 5-point Likert question on the importance of the task. 
Subjective trust in the AI was comparable across conditions (LS: Med.\ = $3.38$, range = [$1.13$--$4.50$]; HS: Med.\ = $3.38$, range = [$2.50$--$4.25$], $p = .23$). While perceived diagnostic importance followed a similar pattern (LS: Med.\ = $5.00$, range = [$4$--$5$]; HS: Med.\ = $5.00$, range = [$2$--$5$], $p = .69$). These results contradict \ref{h3}.

\subsection{Appropriateness of Reliance (AoR)}
\begin{table}[t]
    \centering
    \caption{Descriptive statistics for AoR metrics by condition.}
    \begin{tabular}{l  r r r r r}
    \toprule
    \textbf{Metric} & \textbf{Low} & \textbf{High} & \textbf{$U_{MW}$} & \textbf{$p$} & \textbf{$r$} \\
    \midrule
    RAIR & 0.61 (0.5, 0.8) & 0.67 (0.5, 0.7) & 299 & 0.51 & -0.09 \\
    RSR & 0.42 (0.3, 0.6) & 0.33 (0.3, 0.3) & 440 & 0.054 & 0.27 \\
    \bottomrule
    \end{tabular}
    \label{tab:aor}
\end{table}

To further explore reliance behavior, we analyzed participant RAIR and RSR scores. 
Scores in both metrics did not meet the assumption of normality, so we present medians with 95\% bootstrap confidence intervals and compare conditions with a Mann-Whitney U test. 

Participants from both conditions tended to have higher RAIR and lower RSR scores (\Cref{tab:aor}), indicating that under both conditions, participants relied on  AI recommendations even when the model was incorrect, as shown by the low RSR. Comparison across the two conditions did not result in significant differences for the two metrics. However, participants with raised stakes trended toward a lower median RSR, with a medium effect, than low-stakes participants, suggesting that participants were more likely to defer to incorrect AI under raised stakes.

\subsection{Influence of Decision Time on AoR}
\begin{table}[t]
\renewcommand{\arraystretch}{1}
\centering
\caption{Effects of decision time and stakes on CAIR (posterior estimates, odds ratios ($OR$), 95\% HDIs, and $P(\beta \gtrless 0)$).}

\resizebox{\columnwidth}{!}{
\begin{tabular}{lrrrr}
\toprule

Predictor & Estimate & $OR$ & 95\% HDI & $P(\beta \gtrless 0)$ \\
\midrule
Time & -0.55 & 0.58 & [-1.00, -0.13] & 0.99\,\smarrow{\downarrow} \\
Stakes (high) & 0.32 & 1.37 & [-0.29, 0.95] & 0.83\,\smarrow{\uparrow} \\
Time $\times$ Stakes (high) & 0.37 & 1.45 & [-0.15, 0.93] & 0.91\,\smarrow{\uparrow}\\
\bottomrule
\end{tabular}
}
\label{tab:aor-time:cair}
\end{table}

\begin{table}[t]
\renewcommand{\arraystretch}{1}
\centering
\caption{Effects of decision time and stakes on CSR.}

\resizebox{\columnwidth}{!}{
\begin{tabular}{lrrrr}
\toprule
Predictor & Estimate & $OR$ & 95\% HDI & $P(\beta \gtrless 0)$ \\
\midrule
Time & 1.13 & 3.09 & [0.03, 1.95] & 1.00\,\smarrow{\uparrow} \\
Stakes (high) & -0.74 & 0.48 & [-1.55, 0.11] & 0.95\,\smarrow{\downarrow} \\
Time $\times$ Stakes (high) & -1.13 & 0.32 & [-2.03, -0.25] & 0.99\,\smarrow{\downarrow} \\
\bottomrule
\end{tabular}
}
\label{tab:aor-time:csr}
\end{table}

Given that raised stakes were associated with longer deliberation times and increased overreliance, we further examined this relationship using the AoR framework. 
We modeled individual reliance outcomes (CAIR, CSR) as a function of decision time, stakes, and their interaction using a Bayesian mixed-effects model (Eq.~\ref{eq:fem}) with random effects for participants and Blockies.  We interpret effects to be credible if the 95\% high-density interval (HDI) excludes zero.

Decision time showed opposite effects across AoR outcomes. Longer deliberation was associated with increased underreliance (lower CAIR) (\Cref{tab:aor-time:cair}) and improved self-reliance (higher CSR) (\Cref{tab:aor-time:csr}), with 95\% HDIs excluding zero. In the time $\times$ stakes interaction, however, this pattern shifted, with a trend toward increased CAIR, although the corresponding 95\% HDIs included zero, and reduced CSR, supported by 95\%HDIs excluding zero. This suggests that raised stakes led participants to rely more on the AI even when it was incorrect.

\section{Discussion \& Conclusion}\label{sec:disc}   

This work introduced \blockies, a parametric dataset generator for studying reliance behavior in human-AI decision-making, addressing limitations in existing evaluation approaches. Prior work often relied on domain experts~\cite{amarasinghe_importance_2024,chanda_dermatologist-like_2024}, which is costly and difficult to scale, or on proxy tasks that may not generalize to real-world settings~\cite{bucinca_proxy_2020}. In contrast, \blockies enables the generation of realistic visual diagnostic tasks with fine-grained control over traits, symptoms, and induced biases, making studies with general users more accessible and scalable. The dataset and task are designed to support training with modern AI architectures and to introduce qualities (e.g., interpretability gaps and spatial dependencies) that are challenging for XAI methods, making it well-suited for evaluating real-world XAI interventions for Human-AI decision-making, as human-centered evaluation of XAI is becoming increasingly important~\cite{lai_towards_2023,vilone_explainable_2020,lekadir_future-ai_2025}. 

Using \blockies, we conducted a study to investigate user reliance behavior under raised stakes. Our analysis revealed that participants in the raised-stakes group deliberated longer when supported by AI, suggesting increased effort. However, this did not translate to more calibrated reliance. Instead, raised stakes were associated with a tendency toward increased overreliance. Overall, longer deliberation was associated with greater self-reliance, with participants more likely to keep their own judgments, even when they were incorrect.  However, under raised stakes, this pattern shifted with longer deliberation increasingly associated with reliance on the AI, even when it was incorrect. These findings suggest that increased effort under higher stakes does not necessarily improve reliance calibration, contradicting prior work~\cite{bucinca_trust_2021,vasconcelos_explanations_2023}. One possible explanation is that the complexities of \blockies increase uncertainty, making it more difficult for participants to validate their own judgments. Under pressure, this may lead them to defer to the AI, possibly due to decision inertia or diffusion of responsibility~\cite{sautua_does_2017, anderson_psychology_2006,mei_passing_2026}.

As with any empirical study, limitations should be considered. Using a crowd platform enables scalable evaluation, but participants may not fully reflect expert decision-making processes, limiting generalizability. We aim to mitigate this through application-grounded task design. Additionally, stakes were manipulated at the task level through a performance-based bonus rather than per decision. This may reduce their granularity and differs from utility-based incentive structures of prior work~\cite{Wang2022_}. Despite these limitations, the observed behavioral differences under raised stakes highlight the importance of incorporating realistic consequences in human-AI decision-making studies.  

Overall, our findings suggest that increasing perceived stakes may amplify effort without improving reliance calibration, and may even increase overreliance on incorrect AI advice. Moving forward, we plan to build on this by conducting further studies using \blockies to assess the influence of XAI types, such as feature-, concept-, or example-based explanations~\cite{johnson_explainable_2024}, and XAI-based interventions on reliance behavior and decision effort.

\begin{acks}
This work is associated with the Transregional Collaborative Research Centre (TRR) 318 “Constructing Explainability” and funded by Bielefeld University.

We sincerely thank Olya Hakobyan, Seham Nasr, Hanna Drimalla, and the Human-Centered AI group at Bielefeld University for their valuable support in developing and piloting this work.

We acknowledge the use of large language models (LLMs) for editing and refining the authors’ original content, and for providing support in code implementation. All outputs were reviewed and verified by the authors, who take full responsibility for the final content.
\end{acks}

\bibliographystyle{ACM-Reference-Format}
\bibliography{references}

\end{document}